\documentclass[prd,superscriptaddress,amsfonts,amssymb,amsmath,showpacs,twocolumn]{revtex4-2}
\usepackage{bm}
\usepackage{amsfonts}
\usepackage{latexsym}
\usepackage[latin1]{inputenc}
\usepackage{graphicx}
\usepackage{amsmath}
\usepackage{palatino}
\usepackage{mathpazo}
\usepackage{textcomp}
\usepackage{float}
\usepackage{booktabs}
\usepackage{dcolumn}
\usepackage{hyperref}
\hypersetup{colorlinks,citecolor=blue}
\usepackage{amsmath}
\usepackage{xcolor}
\usepackage{orcidlink}
\usepackage[caption=false]{subfig}
\usepackage{commath}
\def\jnl@style{\it}
\def\aaref@jnl#1{{\jnl@style#1}}

\def\aaref@jnl#1{{\jnl@style#1}}

\def\aj{\aaref@jnl{AJ}}                   % Astronomical Journal
\def\apj{\aaref@jnl{ApJ}}                 % Astrophysical Journal
\def\apjl{\aaref@jnl{ApJ}}                % Astrophysical Journal, Letters
\def\apjs{\aaref@jnl{ApJS}}               % Astrophysical Journal, Supplement
\def\apss{\aaref@jnl{Ap\&SS}}             % Astrophysics and Space Science
\def\aap{\aaref@jnl{A\&A}}                % Astronomy and Astrophysics
\def\aapr{\aaref@jnl{A\&A~Rev.}}          % Astronomy and Astrophysics Reviews
\def\aaps{\aaref@jnl{A\&AS}}              % Astronomy and Astrophysics, Supplement
\def\mnras{\aaref@jnl{Mon.~Not.~Roy.~Astron.~Soc.}}             % Monthly Notices of the RAS
\def\prd{\aaref@jnl{Phys.~Rev.~D}}        % Physical Review D
\def\prc{\aaref@jnl{Phys.~Rev.~C}}  % Physical Review C
\def\prl{\aaref@jnl{Phys.~Rev.~Lett.}}    % Physical Review Letters
\def\qjras{\aaref@jnl{QJRAS}}             % Quarterly Journal of the RAS
\def\skytel{\aaref@jnl{S\&T}}             % Sky and Telescope
\def\ssr{\aaref@jnl{Space~Sci.~Rev.}}     % Space Science Reviews
\def\zap{\aaref@jnl{ZAp}}                 % Zeitschrift fuer Astrophysik
\def\nat{\aaref@jnl{Nature}}              % Nature
\def\aplett{\aaref@jnl{Astrophys.~Lett.}} % Astrophysics Letters
\def\apspr{\aaref@jnl{Astrophys.~Space~Phys.~Res.}} % Astrophysics Space Physics Research
\def\physrep{\aaref@jnl{Phys.~Rep.}}      % Physics Reports
\def\physscr{\aaref@jnl{Phys.~Scr}}       % Physica Scripta
\def\commat{\aaref@jnl{Comm.~Math.~Phys.}}              % Communications in Mathematical Physics
\def\science{\aaref@jnl{Science}}               % Science
\def\cqg{\aaref@jnl{Classical Quant.~Grav.}}            % Classical and Quantum Gravity
\def\jpcs{\aaref@jnl{JPCS}}                                     % Journal of Physics Conference Series
\def\ijmpd{\aaref@jnl{Int.~J.~Mod.~Phys.~D}}                    % International Journal of Modern Physics D
\def\grg{\aaref@jnl{Gen.~Relat.~Gravit.}}               % General Relativity and Gravitation
\def\rpp{\aaref@jnl{Rep.~Prog.~Phys.}}          % Reports on Progress in Physics
\def\npa{\aaref@jnl{Nucl.~Phys.~A}}        % Nuclear Physics A
\def\lrr{\aaref@jnl{Living Rev.~Rel.}}                   % Living reviews in relativity
\def\jcap{\aaref@jnl{J.~Cosmology Astropart.~Phys.}}    % Journal of cosmology and astroparticle physics
\def\rmp{\aaref@jnl{Rev.~Mod.~Phys.}}   %Reviews of modern physics
\def\epjc{\aaref@jnl{Eur.~Phys.~J.~C}}

\allowdisplaybreaks[1]
\begin{document}

\color{black}       %% For one column

\title{Quintessence like behavior of symmetric teleparallel dark energy: Linear and nonlinear model}

\author{A. Hanin}
\email{abdeljalil.hanin.2019@gmail.com}
\affiliation{Quantum Physics and Magnetism Team, LPMC, Faculty of Science Ben
M'sik,\\
Casablanca Hassan II University,
Morocco.}

\author{M. Koussour\orcidlink{0000-0002-4188-0572}}
\email{pr.mouhssine@gmail.com}
\affiliation{Quantum Physics and Magnetism Team, LPMC, Faculty of Science Ben
M'sik,\\
Casablanca Hassan II University,
Morocco.}

\author{Z. Sakhi}
\email{zb.sakhi@gmail.com}
\affiliation{Quantum Physics and Magnetism Team, LPMC, Faculty of Science Ben
M'sik,\\
Casablanca Hassan II University,
Morocco.}
\affiliation{Lab of High Energy Physics, Modeling and Simulations, Faculty of
Science,\\
University Mohammed V-Agdal, Rabat, Morocco.}

\author{M. Bennai\orcidlink{0000-0002-7364-5171}}
\email{ mdbennai@yahoo.fr}
\affiliation{Quantum Physics and Magnetism Team, LPMC, Faculty of Science Ben
M'sik,\\
Casablanca Hassan II University,
Morocco.}
\affiliation{Lab of High Energy Physics, Modeling and Simulations, Faculty of
Science,\\
University Mohammed V-Agdal, Rabat, Morocco.}

%%%%%%%%%%%%%%%%%%%%%%%%%%%%%%%%%%%%  DATE  %%%%%%%%%%%%%%%%%%%%%%%%%%%%%%%%%%%%
\date{\today}
\begin{abstract}

In Einstein's General Relativity (GR), the gravitational interactions are described by the spacetime curvature. Recently, other alternative geometric
formulations and representations of GR have emerged in which the
gravitational interactions are described by the so-called torsion or
non-metricity. Here, we consider the recently proposed modified symmetric
teleparallel theory of gravity or $f\left( Q\right) $ gravity, where $Q$
represents the non-metricity scalar. In this paper, motivated by several
papers in the literature, we assume the power-law form of the function $%
f\left( Q\right) $ as $f\left( Q\right) =\alpha Q^{n+1}+\beta $ (where $%
\alpha $, $\beta $, and $n$ are free model parameters) that contains two
models: Linear ($n=0$) and nonlinear ($n\neq 0$). Further, to add
constraints to the field equations we assume the deceleration parameter form
as a divergence-free parametrization. Then, we discuss the behavior of
various cosmographic and cosmological parameters such as the jerk, snap,
lerk, $Om$ diagnostic, cosmic energy density, isotropic pressure, and
equation of state (EoS) parameter with a check of the violation of the
strong energy condition (SEC) to obtain the acceleration phase of the
Universe. Hence, we conclude that our cosmological $f(Q)$ models behave like
quintessence dark energy (DE).

\end{abstract}

\maketitle

\section{Introduction}

\label{sec1}

As our Universe is in a state of accelerating expansion behavior according
to several observational data, especially from type Ia Supernova (SNIa) (Riess et al. 1998; Perlmutter et al. 1999), Cosmic Microwave Background (CMB) (Caldwell et al. (2004); Huang et al. (2006),
Wilkinson Microwave Anisotropy Probe (WMAP) results (Bennett et al. 2003;  Spergel et al. 2003),
Large-Scale Structures (LSS) (Koivisto and Mota 2006), and Baryonic Acoustic
Oscillations (BAO) (Eisenstein et al. 2005; Percival at el. 2005). So really, we are asking for the help
of several theories to intervene in order to explain the phenomenon of the
observed accelerated expansion. In General Relativity (GR) this phenomenon
is explained as the presence of an unknown form of energy that is behind a
very strong negative pressure called Dark Energy (DE), so to know more one
has to predict the fundamental nature of this DE component. In the
literature, there are many DE models that can be distinguished by the
Equation of State (EoS) parameter $\omega =\frac{p}{\rho }$, where $\rho $
and $p$ represent the cosmic energy density and isotropic pressure,
respectively. Recent cosmological data from SNIa and WMAP indicate that the
present value of the EoS parameter is $\omega _{0}=-1.084\pm 0.063$ and $%
\omega _{0}=-1.073\pm _{0.089}^{0.090}$, respectively. The cosmological
constant $\Lambda $ in GR replaces the prediction of DE and is always given
by an EoS parameter value $\omega _{\Lambda }=-1$. In other words, there is
no compatibility between this cosmological constant value that acquired by
the quantum gravity model and the value obtained by observation (Weinberg 1989).
This phenomenon is called the cosmological constant problem. Since the
evolution of the two is so different, i.e. between the matter density and DE
density, they have almost the same order of magnitude, which this observed
coincidence between the densities is called the cosmic coincidence problem.
Among many interesting models of DE, there is a very important one called
quintessence DE which focuses on the scalar field in the $\omega $ range $%
-1<\omega <-\frac{1}{3}$. In parallel with the quintessence idea, the DE
density decreases with the fourth dimension called cosmic time (Ratra and Peebles 1998; Xu et al. 2011)%
. Moreover, it is still very sensitive to predict exactly what DE nature?,
another proposition called phantom energy which is determined by the value $%
\omega <-1$ which remains the core of the unknown for all researchers now,
rather behaving strangely. However, when we say phantom energy, the DE
density increases with cosmic time, which implies the existence of problems
preventing our scientific research path and decelerating our speed to
understand our Universe, here we are talking about the finite-time future
singularity (Caldwell et al. 2003; Barrow 2004). Thus, from the above discussion, the
observational value of the EoS parameter favours the phantom and
quintessence DE models.

To attack the problem of the late-time accelerated expansion of the
Universe, modified gravitational theories (MGT) have been suggested in the
literature as an auxiliary alternative to Einstein's gravitational theory.
In our current analysis, we will study different mechanisms of DE in the $%
f(Q)$ gravity model, where $Q$ is the non-metricity scalar responsible for
gravity. An interesting symmetric teleparallel gravity (or $f\left( Q\right) 
$ gravity), which was developed for the first time in (Jimenez et al. 2018; 2020).
In Weyl's geometry, the covariant derivative of the metric tensor is non
null and this characteristic can be displayed mathematically in terms of a
novel geometric quantity, named non-metricity (Xu et al. 2019). Geometrically, the
non-metricity can be described as the variation of the length of a vector
during parallel transport. For a better understanding of our Universe, we
are obligated to replace the curvature concept with a more general
geometrical concept. Among the most important geometrical tools that can
successfully demonstrate gravity, there are two equivalents: the torsion $T$
and non-metricity $Q$ representations. The first representation has a zero
curvature and non-metricity with a spacetime torsion, famous as the \textit{%
Teleparallel Equivalent of GR} (TEGR), In contrast, the second
representation uses zero curvature and torsion with the presence of
spacetime non-metricity, called the \textit{Symmetric Teleparallel
Equivalent of GR} (STEGR). In the TEGR representation, the metric tensor $%
g_{\mu \nu }$ replaced by the set of tetrad vectors $e_{\mu }^{i}$. The
torsion, created by the tetrad fields, can then be exploited to entirely
explain gravitational effects (Xu et al. 2019). The generalized version of GR gives
rise to the $f(R)$ theory of gravity in which the gravitational effect is
related to the non-zero curvature with zero torsion and non-metricity (Starobinsky 1979), while the generalization of the TEGR version is named $f(T)$
gravity, in which spacetime is defined by a non-zero torsion with zero
curvature and non-metricity (Bengochea and Ferraro 2009). Also, the $f(Q)$ theory is a
generalized version of the STEGR in which the non-metricity scalar explains
the gravitational effects with zero curvature and torsion. The $f(Q)$
gravity has been investigated from various angles such as energy conditions 
(Mandal et al. 2020), covariant formulation (Zhao 2022), cosmography (Mandal et al. 2020),
signature of $f(Q)$ theory (Frusciante 2021) and anisotropic nature of spacetime (Koussour et al. 2022; Koussour and Bennai 2022; Koussour et al. 2022).

Our work is structured as follows: In Sec. \ref{sec2}, we introduce the some
basics of $f\left( Q\right) $ gravity theory. In Sec. \ref{sec3}, we propose
a cosmological $f(Q)$ gravity model and derive the various cosmographic and
cosmological parameters such as the jerk, snap, lerk, $Om$ diagnostic,
cosmic energy density, isotropic pressure, and equation of state (EoS)
parameter. Then, we discuss the different energy conditions of our
cosmological model in Sec. \ref{sec4}. F Finally, we present our conclusions
in Sec \ref{sec5}.

\section{Some basics of $f(Q)$ gravity theory}

\label{sec2}

It is known in differential geometry that the general connection $\Sigma
_{\mu \nu }^{\alpha }$ helps us in the parallel transport of the vectors and
the notion of covariant derivatives, while the so-called metric tensor $%
g_{\mu \nu }$ helps us to determine angles, volumes, distances, etc. It is a
generalization of the so-called gravitational potential of the classical
theory. Generally, this general connection $\Sigma _{\mu \nu }^{\alpha }$
can be decomposed into all possible contributions (i.e. in the presence of
torsion $T$ and non-metricity $Q$ terms next to the curvature $R$) as (Ortin 2015),%
\begin{equation}
\Sigma _{\ \mu \nu }^{\alpha }=\Gamma _{\ \mu \nu }^{\alpha }+C_{\ \mu \nu
}^{\alpha }+L_{\ \mu \nu }^{\alpha },  \label{2a}
\end{equation}%
with the famous Levi-Civita connection $\Gamma _{\ \mu \nu }^{\alpha }$ of
the metric tensor $g_{\mu \nu }$ is,%
\begin{equation}
\Gamma _{\ \mu \nu }^{\alpha }\equiv \frac{1}{2}g^{\alpha \lambda }(g_{\mu
\lambda ,\nu }+g_{\lambda \nu ,\mu }-g_{\mu \nu ,\lambda }),  \label{2b}
\end{equation}%
and the expression for the \textit{Contortion tensor} $C_{\ \mu \nu
}^{\alpha }$\ is,%
\begin{equation}
C_{\ \mu \nu }^{\alpha }\equiv \frac{1}{2}(T_{\ \mu \nu }^{\alpha }+T_{\mu \
\nu }^{\ \alpha }+T_{\nu \ \mu }^{\ \alpha })=-C_{\ \nu \mu }^{\alpha }.
\label{2c}
\end{equation}

Finally, the \textit{Disformation tensor} $L_{\ \mu \nu }^{\alpha }$\ is
given\ as,%
\begin{equation}
L_{\ \mu \nu }^{\alpha }\equiv \frac{1}{2}(Q_{\ \mu \nu }^{\alpha }-Q_{\mu \
\nu }^{\ \alpha }-Q_{\nu \ \mu }^{\ \alpha })=L_{\ \nu \mu }^{\alpha }.
\label{2d}
\end{equation}

For $T_{\ \mu \nu }^{\alpha }$ and $Q_{\alpha \mu \nu }$ terms in Eqs. %
\eqref{2c} and \eqref{2d}, are famous as the torsion tensor and the
non-metricity tensor, respectively. Its expression is given as,%
\begin{equation}
T_{\ \mu \nu }^{\alpha }\equiv \Sigma _{\ \mu \nu }^{\alpha }-\Sigma _{\ \nu
\mu }^{\alpha },  \label{2e}
\end{equation}%
and%
\begin{equation}
Q_{\alpha \mu \nu }\equiv \nabla _{\alpha }g_{\mu \nu }\neq 0.  \label{2f}
\end{equation}

As we mentioned above, the connection presumed to be the torsion and
curvature vanish within the so-called \textit{Symmetric Teleparallel
Equivalent to General Relativity} (STEGR), such that it conforms to a pure
coordinate transformation of trivial connection as shown in (Jimenez et al. 2018).
The components of the connection in Eq. (\ref{2a}) can be rewritten as,%
\begin{equation}
\Sigma ^{\alpha }\,_{\mu \beta }=\frac{\partial y^{\alpha }}{\partial \xi
^{\rho }}\partial _{\mu }\partial _{\beta }\xi ^{\rho }.  \label{2g}
\end{equation}

In the above equation, $\xi ^{\alpha }=\xi ^{\alpha }(y^{\mu })$ is an
invertible relation and $\frac{\partial y^{\alpha }}{\partial \xi ^{\rho }}$
is the inverse of the corresponding Jacobian (Jimenez et al. 2020). It is constantly
feasible to get a coordinate system in which the connection $\Sigma _{\ \mu
\nu }^{\alpha }$ becomes zero (Adak et al. 2013). This situation is called
coincident gauge (Adak et al. 2006; Mol 2017). Hence, in this choice, the covariant
derivative $\nabla _{\alpha }$ reduces to the partial derivative $\partial
_{\alpha }$ i.e. $Q_{\alpha \mu \nu }=\partial _{\alpha }g_{\mu \nu }$.
However, it is clear from the previous discussion that the Levi-Civita
connection $\Gamma _{\ \mu \nu }^{\alpha }$ can be written in terms of the
disformation tensor $L_{\ \mu \nu }^{\alpha }$ as $\Gamma _{\ \mu \nu
}^{\alpha }=-L_{\ \mu \nu }^{\alpha }$.

The action that conforms with STEGR is described by%
\begin{equation}
S_{STEGR}=\int {\frac{1}{2}}\left( -{Q}\right) {\sqrt{-g}d^{4}x}+\int {L_{m}%
\sqrt{-g}d^{4}x.}
\end{equation}%
where $g$ being the determinant of the tensor metric $g_{\mu \nu }$ and ${%
L_{m}}$ the matter lagrangian density. Moreover, throughout this article we
will consider natural units. The modified $f\left( R\right) $\ gravity is a
generalization of GR, and the $f\left( T\right) $ gravity is a
generalization of TEGR. Thus, in the same way, the $f\left( Q\right) $ is a
generalization of STEGR in which the extended action is given by, 
\begin{equation}
S=\int {\frac{1}{2}f(Q)\sqrt{-g}d^{4}x}+\int {L_{m}\sqrt{-g}d^{4}x.}
\label{2j}
\end{equation}

Here $f(Q)$ is an arbitrary function of the non-metricity scalar $Q$. The
so-called STEGR can be obtained by assuming the following functional form $%
f\left( Q\right) =-Q$, see Ref. (Jimenez et al. 2018). In addition, the non-metricity
tensor in Eq. (\ref{2f}) has the following two independent traces,%
\begin{equation}
Q_{\alpha }=Q_{\alpha }{}^{\mu }{}_{\mu }\text{ \ \ and \ \ }\tilde{Q}%
_{\alpha }=Q^{\mu }{}_{\alpha \mu }.  \label{2k}
\end{equation}

Further, the non-metricity conjugate (superpotential tensor) is given by,%
\begin{equation}
4P^{\lambda }{}_{\mu \nu }=-Q^{\lambda }{}_{\mu \nu }+2Q_{(\mu }{}^{\lambda
}{}_{\nu )}+(Q^{\lambda }-\tilde{Q}^{\lambda })g_{\mu \nu }-\delta _{(\mu
}^{\lambda }Q_{\nu )}.  \label{2l}
\end{equation}

The non-metricity scalar can be acquired as, 
\begin{equation}
Q=-Q_{\lambda \mu \nu }P^{\lambda \mu \nu }.  \label{2m}
\end{equation}

Now, the energy-momentum tensor of the content of the Universe as a perfect
fluid matter is given as,%
\begin{equation}
T_{\mu \nu }=\frac{-2}{\sqrt{-g}}\frac{\delta (\sqrt{-g}L_{m})}{\delta
g^{\mu \nu }}.  \label{2n}
\end{equation}

By varying the above action \eqref{2j} with regard to the metric tensor $%
g_{\mu \nu }$\ components yield 
\begin{widetext}
\begin{equation}\label{2p}
\frac{2}{\sqrt{-g}}\nabla_\lambda (\sqrt{-g}f_Q P^\lambda\:_{\mu\nu}) + \frac{1}{2}g_{\mu\nu}f+f_Q(P_{\mu\lambda\beta}Q_\nu\:^{\lambda\beta} - 2Q_{\lambda\beta\mu}P^{\lambda\beta}\:_\nu) = -T_{\mu\nu}.
\end{equation}
\end{widetext}

Here, for simplicity we consider $f_{Q}=\frac{df}{dQ}$ . Again, by varying
the action with regard to the connection, we can get as a result, 
\begin{equation}
\nabla _{\mu }\nabla _{\nu }(\sqrt{-g}f_{Q}P^{\mu \nu }:_{\lambda })=0.
\label{2q}
\end{equation}

According to recent observations of the CMB, our Universe is homogeneous and
isotropic on a large scale, that is to say on a scale more significant than
the scale of galaxy clusters. For this, in our current analysis, we consider
a flat FLRW background geometry in Cartesian coordinates with a metric,%
\begin{equation}
ds^{2}=-dt^{2}+a^{2}(t)[dx^{2}+dy^{2}+dz^{2}],  \label{3a}
\end{equation}%
where $a(t)$ is the scale factor of the Unverse. Furthermore, the
non-metricity scalar corresponding to the metric (\eqref{3a}) is obtained as%
\begin{equation}
Q=6H^{2},  \label{3b}
\end{equation}%
where $H$ is the Hubble parameter that measures the rate of expansion of the
Universe.

In cosmology, the most commonly used energy-momentum tensor is the perfect
cosmic fluid, i.e. without considering viscosity effects. In this case,%
\begin{equation}
T_{\mu \nu }=(\rho +p)u_{\mu }u_{\nu }+pg_{\mu \nu },  \label{3c}
\end{equation}%
where $\rho $ and $p$\ represent the cosmic energy density and isotropic
pressure of the perfect cosmic fluid respectively, and $u^{\mu }=(1,0,0,0)$
represents the four velocity vector components characterizing the fluid.%
\newline

The modified Friedmann equations that describe the dynamics of the Universe
in $f\left( Q\right) $ gravity are (Lazkoz et al. 2019; Harko et al. 2018)%
\begin{equation}
3H^{2}=\frac{1}{2f_{Q}}\left( -\rho +\frac{f}{2}\right) ,  \label{F1}
\end{equation}%
and%
\begin{equation}
\dot{H}+3H^{2}+\frac{\dot{f}_{Q}}{f_{Q}}H=\frac{1}{2f_{Q}}\left( p+\frac{f}{2%
}\right) ,  \label{F2}
\end{equation}%
where an overhead dot points out the differentiation of the quantity with
regard to cosmic time $t$. Also, it is good to point out that the standard
Friedmann equations of General Relativity are obtained if the function $%
f(Q)=-Q$ is assumed (Lazkoz et al. 2019).

Now, we get the matter/energy conservation equation in its famous form as, 
\begin{equation}
\dot{\rho}+3H\left( \rho +p\right) =0  \label{3f}
\end{equation}

Using Eqs. (\ref{F1}) and (\ref{F2}), the expressions for the cosmic energy
density $\rho $ and isotropic pressure $p$ of the fluid are 
\begin{equation}
\rho =\frac{f}{2}-6H^{2}f_{Q},  \label{F22}
\end{equation}%
\begin{equation}
p=\left( \dot{H}+3H^{2}+\frac{\dot{f_{Q}}}{f_{Q}}H\right) 2f_{Q}-\frac{f}{2}.
\label{F33}
\end{equation}

Furthermore, by using Eqs. (\ref{F1}) and (\ref{F2}) we can get the field
equations like the standard Friedmann equations in GR, by inserting the
concept of an effective cosmic energy density $\overline{\rho }$ and an
effective isotropic pressure $\overline{p}$ as%
\begin{equation}
3H^{2}=-\frac{1}{2}\overline{\rho }=-\frac{1}{2f_{Q}}\left( \rho -\frac{f}{2}%
\right) ,  \label{F222}
\end{equation}%
\begin{equation}
\dot{H}+3H^{2}=\frac{\overline{p}}{2}\,=-\frac{\dot{f_{Q}}}{f_{Q}}H+\frac{1}{%
2f_{Q}}\left( p+\frac{f}{2}\right) .  \label{F333}
\end{equation}

\section{Cosmological $f(Q)$ Model}

\label{sec3}

Motivated by the work of Capozziello et al. (2022)  where it was
found that the top approximation for characterizing the accelerated
expansion of the Universe in $f(Q)$ gravity is constituted by a scenario
with $f(Q)=\alpha +\beta Q^{n}$, in our current analysis, we examine a
power-law form of function $f\left( Q\right) $ given by $f(Q)=\alpha
Q^{n+1}+\beta $, where $\alpha $, $\beta $ and $n$ are free model
parameters. It is important to mention that $n=0$ corresponds to $%
f(Q)=\alpha Q+\beta $, i.e. a case of the linear model, while $n\neq 0$
corresponds to a case of the nonlinear model. The value of $f_{Q}$ in the
field equations (\ref{F1}) and (\ref{F2}) is obtained as $f_{Q}=\alpha
\left( n+1\right) Q^{n}$.

\subsection{Cosmographic Parameters}

Looking at the modified Friedmann equations (\ref{F1}) and (\ref{F2}), they
are two differential equations with three unknowns $\rho $, $p$, and $H$ ($%
Q=6H^{2}$). Thus, in the present scenario, to find the exact solutions of
these two field equations, we need one more additional equation. This
additional equation is a parametrization of the Hubble parameter in general.
However, here we are concerned with the study of the accelerating expansion
of the Universe, we will consider an additional equation for the
deceleration parameter as a divergence-free parametrization (Mamon and Das 2016; Hanafy and Nashed 2019; Gadbail et al. 2022),%
\begin{equation}
q\left( z\right) =q_{0}+q_{1}\frac{z\left( 1+z\right) }{1+z^{2}},  \label{qz}
\end{equation}%
where $q_{0}$ is the current value of $q\left( z\right) $ and $q_{1}$
constitutes the variation of the deceleration parameter with respect to $z$.
Furthermore, these two parameters $q_{0}$ and $q_{1}$ are gained from the
observational constraints. According to Ref. (Mamon and Das 2016), the above
parametrization reduces to $q\left( z\right) =q_{0}+q_{1}$ at high redshift
(i.e. $z>>1$), while it reduces to linear expansion form i.e. $q\left(
z\right) =q_{0}+q_{1}z$ at low redshift (i.e. $z<<1$). The main motivation
for this choice is that it provides a finite value for the deceleration
parameter in the entire range i.e. $z\in \left[ -1,\infty \right] $. It is
therefore valid for the entire evolutionary history of the Universe. Also,
it is useful to mention here that the assumed parametric form of $q\left(
z\right) $ is inspired by one of the most common divergence-free
parametrizations of the DE equation of state (Barboza and Alcaniz 2008). It appears to
be adaptable enough to match the $q\left( z\right) $ behavior of a large
class of DE models.

To get the expression for the Hubble parameter, we use the following
relation,%
\begin{equation}
H\left( z\right) =H_{0}\exp \left( \int_{0}^{z}\frac{1+q\left( z\right) }{%
\left( 1+z\right) }dz\right) ,  \label{H}
\end{equation}%
which is valid for all parametrizations, and $H_{0}$ is the value of the
Hubble parameter at present.

Introducing (\ref{H}) in (\ref{qz}), we get the expression for the Hubble
parameter in terms of $z$ as,%
\begin{equation}
H\left( z\right) =H_{0}\left( 1+z\right) ^{1+q_{0}}\left( 1+z^{2}\right) ^{%
\frac{q_{1}}{2}}.  \label{Hz}
\end{equation}

In addition, the derivative of the Hubble parameter with respect to cosmic
time can be written in terms of the deceleration parameter as (Mandal et al. 2020),%
\begin{equation}
\overset{.}{H}=-H^{2}\left( 1+q\right)
\end{equation}

In cosmology, to explain the evolution of the Universe it is useful to study
the behavior of the deceleration parameter defined by Eq. (\ref{qz}). It is
an effective tool for knowing the nature of the expansion of the Universe,
i.e. accelerating expansion ($q<0$) or decelerating expansion ($q>0$).
According to recent observations, the Universe is definitely accelerating,
and the current value of $q\left( z\right) $ is negative i.e. $-1\leq q<0$.
So to plot the proposed $q\left( z\right) $ behavior, the two parameters $%
q_{0}$ and $q_{1}$ must be constrained with the observational data.
According to Ref. Gadbail et al. (2022), the observational constraints on the model
parameters were studied using the Bayesian analysis for the OHD
(Observational Hubble Data) and the Pantheon sample (SNeIa). The best fit
values of the model parameters are $q_{0}=-0.611_{-0.051}^{+0.051}$ and $%
q_{1}=0.66_{-0.087}^{+0.087}$\ corresponding to the OHD+SNeIa datasets,
respectively. 
\begin{figure}[tbp]
\includegraphics[scale=0.65]{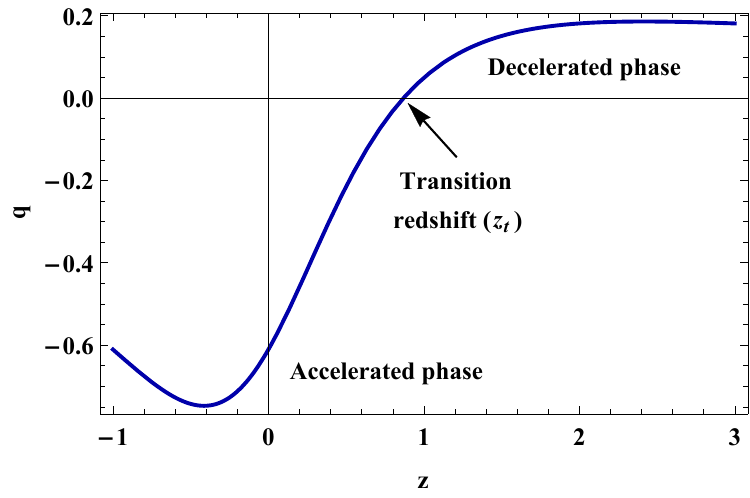}
\caption{Behavior of the deceleration parameter vs redshift.}
\label{Fq}
\end{figure}

The behavior of the deceleration parameter $q$ versus $z$ corresponding to
the values of the parameters constrained by the OHD+SNeIa datasets is
displayed in Fig. \ref{Fq}. It can be seen that $q$ changes from negative to
positive values at the value of the transition redshift $z_{t}=0.87$ with
the current value of $q$ for our model is $q_{0}=-0.611$. Thus, the proposed
model represents the transition of the Universe from the early deceleration
phase to the current acceleration phase as shown by recent observations data 
(Planck 2020). Further, it is good to expand on the discussion of more
geometrical parameters of the DE models to provide essential information
about the evolution of our Universe. In the Taylor series expansion of the
scale factor of the Universe with regard to cosmic time, there come the
derivatives in the highest order of the deceleration parameter, which is
famous as a jerk ($j$), snap ($s$), lerk ($l$) parameters. It can be said
that the jerk parameter represents the evolution of the deceleration
parameter. As we know that we can constrain $q$ from the observations data,
the jerk parameter is studied to predict the future of the Universe and to
compare other DE models with the $\Lambda $CDM model in which the value of $%
j $ is one always. In addition, the jerk parameter along with higher
derivatives such as snap and lerk parameters provide a helpful understanding
of the emergence of sudden future singularities (Pan et al. 2018).

The jerk ($j$), snap ($s$) and lerk ($l$) parameters are defined as (Pan et al. 2018),%
\begin{equation}
j\left( z\right) =\frac{1}{H^{3}}\frac{da^{\left( 3\right) }}{a}=\left(
1+z\right) \frac{dq}{dz}+q\left( 1+2q\right) ,
\end{equation}%
\begin{equation}
s\left( z\right) =\frac{1}{H^{4}}\frac{da^{\left( 4\right) }}{a}=-\left(
1+z\right) \frac{dj}{dz}-j\left( 2+3q\right) ,
\end{equation}%
\begin{equation}
l\left( z\right) =\frac{1}{H^{5}}\frac{da^{\left( 5\right) }}{a}=-\left(
1+z\right) \frac{ds}{dz}-s\left( 3+4q\right) .
\end{equation}

From the considered $q$ in Eq. (\ref{qz}), the jerk, snap and lerk equations
are given as,%
\begin{widetext}
\begin{equation}
j\left( z\right) =\frac{%
\begin{array}{c}
2q_{0}^{2}\left( z^{2}+1\right) ^{2}+q_{0}\left( z^{2}+1\right) \left(
4q_{1}(z+1)z+z^{2}+1\right) 
+q_{1}(z+1)\left( (2q_{1}+1)z^{3}+(2q_{1}-1)z^{2}+3z+1\right)%
\end{array}%
}{\left( z^{2}+1\right) ^{2}},
\end{equation}
\begin{equation}
s\left( z\right) =-\frac{%
\begin{array}{c}
6q_{0}^{3}\left( z^{2}+1\right) ^{3}+q_{0}^{2}\left( z^{2}+1\right)
^{2}\times \left( 18q_{1}z(z+1)+7\left( z^{2}+1\right) \right)  \\ 
+q_{0}\left( z^{2}+1\right) \times \left(
18q_{1}^{2}z^{2}(z+1)^{2}+7q_{1}\left( 2z^{4}+z^{3}+3z^{2}+5z+1\right)
+2\left( z^{2}+1\right) ^{2}\right)  \\ 
+q_{1}(z+1)\times \left( \left( 6q_{1}^{2}+7q_{1}+2\right) z^{5}+2\left(
6q_{1}^{2}-1\right) z^{4}+2\left( 3q_{1}^{2}+7q_{1}+4\right)
z^{3}+4(7q_{1}-3)z^{2}+(7q_{1}+6)z+6\right) 
\end{array}%
}{\left( z^{2}+1\right) ^{3}},
\end{equation}%
and, 
\begin{equation}
l\left( z\right) =\frac{%
\begin{array}{c}
24q_{0}^{4}\left( z^{2}+1\right) ^{4}+2q_{0}^{3}\left( z^{2}+1\right)
^{3}\left( 48q_{1}z(z+1)+23\left( z^{2}+1\right) \right)  \\ 
+q_{0}^{2}\left( z^{2}+1\right) ^{2}\times \left(
144q_{1}^{2}z^{2}(z+1)^{2}+46q_{1}\left( 3z^{4}+2z^{3}+4z^{2}+6z+1\right)
+29\left( z^{2}+1\right) ^{2}\right)  \\ 
+2q_{0}\left( z^{2}+1\right) \times \left( 
\begin{array}{c}
48q_{1}^{3}z^{3}(z+1)^{3}+23q_{1}^{2}z\left( 3z^{3}-2z^{2}+7z+2\right)
(z+1)^{2} \\ 
+q_{1}\left( 29z^{6}+11z^{5}+76z^{4}+28z^{3}-z^{2}+105z+40\right) +3\left(
z^{2}+1\right) ^{3}%
\end{array}%
\right)  \\ 
+q_{1}(z+1)\times \left( 
\begin{array}{c}
24q_{1}^{3}(z+1)^{3}z^{4}+46q_{1}^{2}(z+1)^{2}\left( z^{3}-z^{2}+3z+1\right)
z^{2} \\ 
+q_{1}\left( 29z^{7}-7z^{6}+101z^{5}-23z^{4}-45z^{3}+223z^{2}+115z+7\right) 
\\ 
+6\left( z^{7}-z^{6}+5z^{5}-5z^{4}+15z^{3}-15z^{2}-5z+5\right) 
\end{array}%
\right) 
\end{array}%
}{\left( z^{2}+1\right) ^{4}}.
\end{equation}
\end{widetext}

\begin{figure}[tbp]
\includegraphics[scale=0.6]{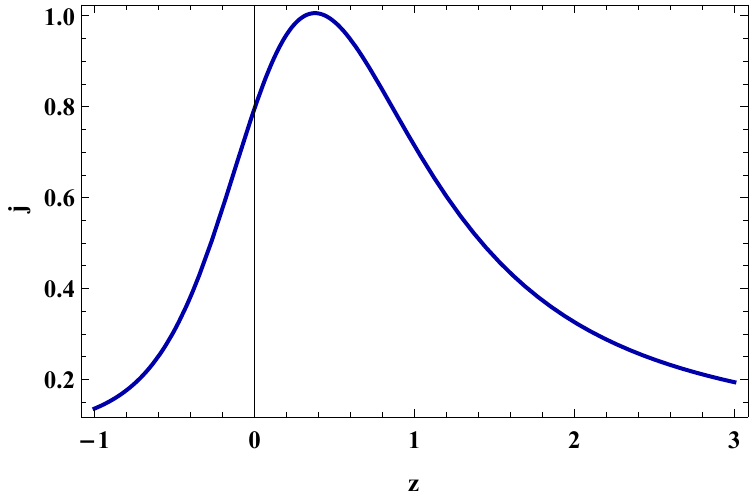}
\caption{Behavior of the jerk parameter vs redshift.}
\label{Fj}
\end{figure}

\begin{figure}[tbp]
\includegraphics[scale=0.6]{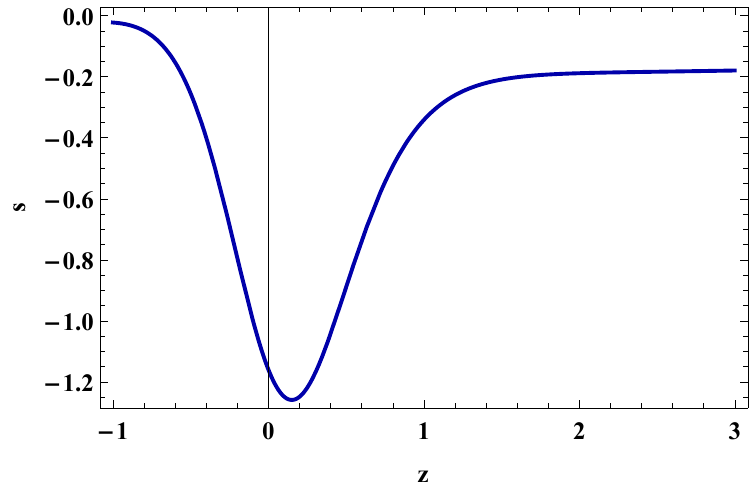}
\caption{Behavior of the snap parameter vs redshift.}
\label{Fs}
\end{figure}

\begin{figure}[tbp]
\includegraphics[scale=0.6]{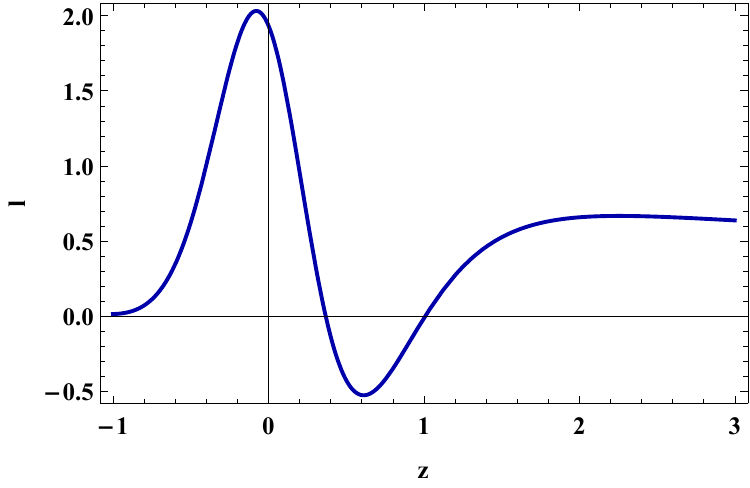}
\caption{Behavior of the lerk parameter vs redshift.}
\label{Fl}
\end{figure}

From Figs. \ref{Fj} and \ref{Fl}, it is clear that the current values of
both parameters: jerk and lerk are positive (i.e. at redshift $z=0$) which
represents an accelerated expansion of the Universe. Fig. \ref{Fs} indicates
the negative behavior of the snap parameter at present ($z=0$) that makes
the Universe in an accelerating expansion currently. Also, the value of the
current jerk parameter is not equal to one, which leads to our model not
being similar to the behavior of the $\Lambda $CDM model at present (at $z=0$%
). Interestingly, this means that under certain modified gravity, late-time
acceleration of the Universe can be seen using geometrical methods.

\subsection{Om Diagnostic}

With the development of modern cosmology and the emergence of many proposed
models to solve the mystery of DE, it has become difficult to differentiate
between these models. For this, an effective diagnostic parameter tool
called $Om$\ has been proposed in this context, extracted from the Hubble
parameter (Sahni et al. 2008). This parameter easily tells us about the dynamical
nature of DE models from the $Om(z)$ slope. The positive slope values of
this diagnostic parameter tool indicate phantom nature ($\omega <-1$), while
its negative values correspond to the quintessence nature ($\omega >-1$).
The diagnostic $Om$ is defined as,%
\begin{equation}
Om\left( z\right) =\frac{\left( \frac{H\left( z\right) }{H_{0}}\right) ^{2}-1%
}{z^{3}+3z^{2}+3z}.
\end{equation}

The diagnostic $Om$ for our current analysis is,%
\begin{equation}
Om\left( z\right) =\frac{(z+1)^{2q_{0}+2}\left( z^{2}+1\right) ^{q1}-1}{%
(z+1)^{3}-1}.
\end{equation}

\begin{figure}[tbp]
\includegraphics[scale=0.6]{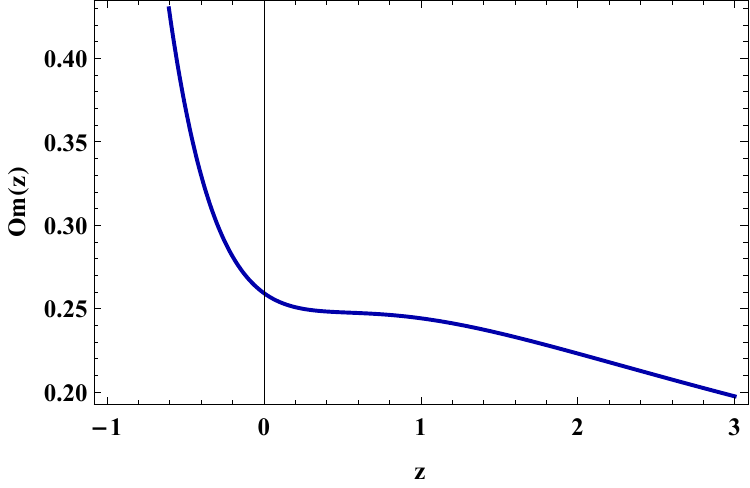}
\caption{Behavior of the $Om\left( z\right) $ parameter vs redshift.}
\label{Fom}
\end{figure}

From Fig. \ref{Fom} it is very clear that the positive slope value of $Om(z)$
indicates a quintessence-like behavior that represents the recently observed
accelerating expansion of the Universe. 

\subsection{Cosmological Parameters}

The evolution of the cosmic energy density, isotropic pressure, and EoS
(Equation of State) parameter in terms of redshift are discussed below. Now,
by using the values of $\rho $ and $p$ from (\ref{F22}) and (\ref{F33}) with
the choice of $f\left( Q\right) $, we obtain the expression for cosmic
energy density $\rho $ and isotropic pressure $p$ as follows:%
\begin{widetext}
\begin{equation}
\rho =\frac{\beta }{2}-\frac{\alpha }{2}6^{n+1}(2n+1)\left( H^{2}\right)
^{n+1},  \label{FF1}
\end{equation}%
\begin{equation}
p=\frac{\alpha }{2}\left( -2^{n+1}\right) 3^{n}(2n+1)\left( H^{2}\right)
^{n+1}(2n(q+1)+2q-1)-\frac{\beta }{2}.  \label{FF2}
\end{equation}
\end{widetext}

\begin{figure}[tbp]
\includegraphics[scale=0.6]{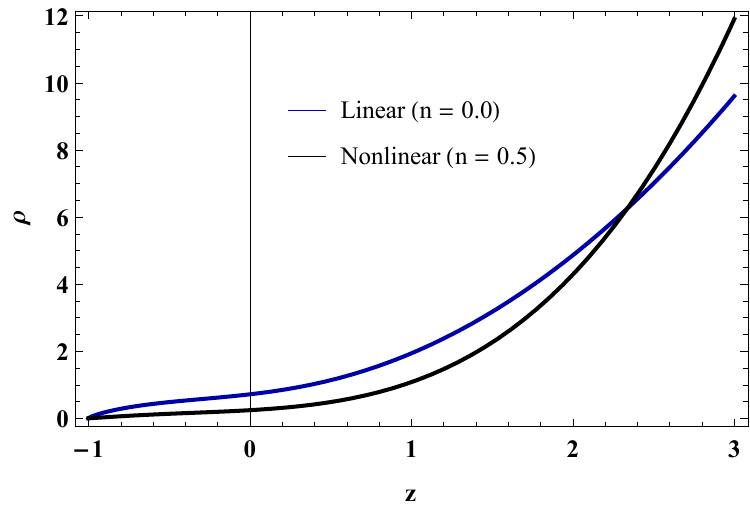}
\caption{Behavior of the cosmic energy density vs redshift.}
\label{Frho}
\end{figure}

\begin{figure}[tbp]
\includegraphics[scale=0.6]{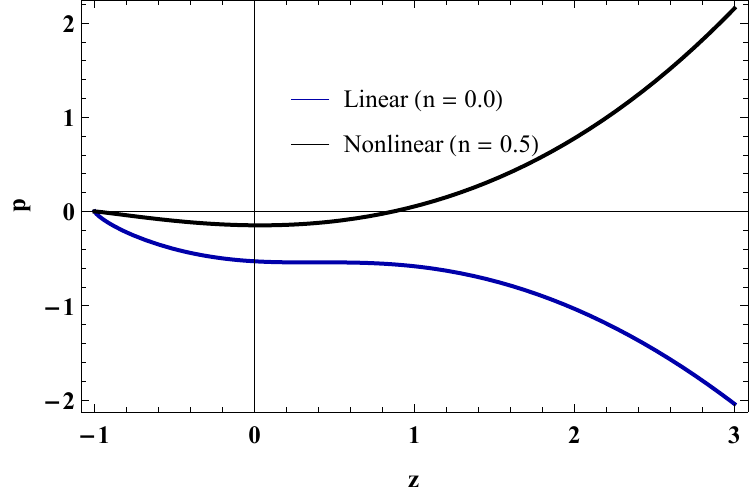}
\caption{Behavior of the isotropic pressure vs redshift.}
\label{Fp}
\end{figure}

The plots in Figs. \ref{Frho} and \ref{Fp}\ exhibit very evident that
the redshift evolution of cosmic energy density and isotropic pressure,
derived here for the FLRW metric in the framework of $f(Q)$ gravity is fully
consistent with the results derived in several works of literature (Shekh 2021, 2021a, 2022, Raja 2021). Specifically, the cosmic energy density is a
positive and increasing function in terms of the redshift of both models
(linear and nonlinear), while the isotropic pressure is negative in the
present and the future. Thus, negative pressure is responsible for the
acceleration phase of the Universe at the present.

The EoS parameter expression ($\omega =\frac{p}{\rho }$) for the cosmic
fluid corresponding to the above cosmic energy density and isotropic
pressure is,%
\begin{equation}
\omega =\frac{\beta +\alpha 2^{n+1}3^{n}(2n+1)\left( H^{2}\right)
^{n+1}(2n(q+1)+2q-1)}{\alpha 6^{n+1}(2n+1)\left( H^{2}\right) ^{n+1}-\beta }.
\end{equation}

\begin{figure}[tbp]
\includegraphics[scale=0.65]{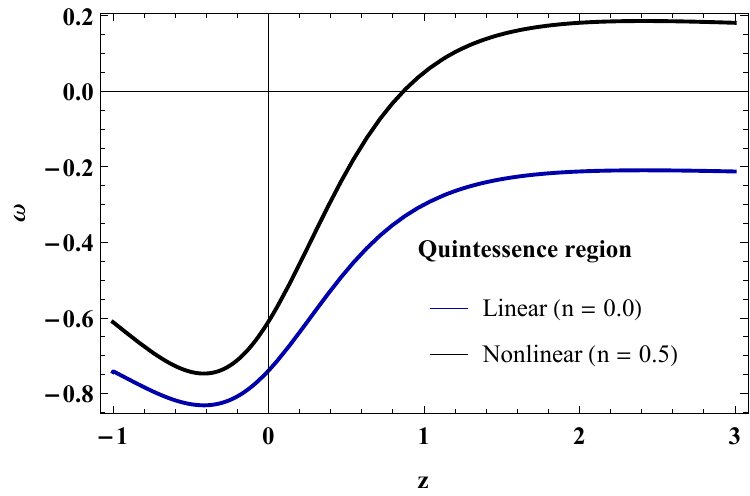}
\caption{Behavior of the EoS parameter vs redshift.}
\label{Fomega}
\end{figure}

In GR, an exotic form of energy (dark energy), described as negative
isotropic pressure ($p<0$) or equivalently with a negative EoS parameter ($%
\omega <0$), has been proposed to be responsible for the accelerating
expansion of the Universe. This last parameter relates to the cosmic energy
density $\rho $ and isotropic pressure $p$ and describes the stages of the
expansion of the Universe. It represents non-relativistic particles (matter)
when $\omega =0$, and if $\omega =\frac{1}{3}$, this describes the
relativistic particles (radiation) phase. The phantom region is represented
by $\omega <-1$, while $-1<\omega \leq -\frac{1}{3}$ exhibits the
quintessence region. Finally, for $\omega =-1$, the behavior of the $\Lambda 
$CDM model is shown. In general, for our model to predict the accelerating
phase of the Universe, it must be $\omega \leq -\frac{1}{3}$. From Fig. \ref%
{Fomega}, it is very clear that our model predicts accelerated expansion,
and behaves like the quintessence model of dark energy in both models.

\section{Energy Conditions}

\label{sec4}

The so-called energy conditions (ECs) play an important role in the
geodesics description of the Universe, and can be obtained from the the
famous Raychaudhuri equation (Raychaudhuri 1955). In addition, ECs can be used to
predict an accelerating Universe by violating the strong energy condition
(SEC). If we consider that the content of the Universe is an perfect fluid,
the ECs in $f(Q)$\ gravity are

\begin{itemize}
\item Weak energy condition (WEC): $\rho +p\geq 0$ and $\rho \geq 0$;

\item Null energy condition (NEC): $\rho +p\geq 0$;

\item Dominant energy condition (DEC): $\rho \geq \left\vert p\right\vert $
and $\rho \geq 0$;

\item Strong energy condition (SEC): $\rho +3p-6\overset{.}{f}_{Q}H+f\geq 0$
\end{itemize}

Taking Eqs. (\ref{FF1}) and (\ref{FF2}) into WEC, NEC, DEC and SEC
conditions, we get

\begin{widetext}
\begin{equation}
\text{WEC}\Leftrightarrow \rho +p=\alpha \left( -2^{n+1}\right) 3^{n}\left(
2n^{2}+3n+1\right) (q+1)\left( H^{2}\right) ^{n+1}\geq 0\text{ and }\rho =%
\frac{\beta }{2}-\frac{\alpha }{2}6^{n+1}(2n+1)\left( H^{2}\right)
^{n+1}\geq 0,
\end{equation}%
\begin{equation}
\text{NEC}\Leftrightarrow \rho +p=\alpha \left( -2^{n+1}\right) 3^{n}\left(
2n^{2}+3n+1\right) (q+1)\left( H^{2}\right) ^{n+1}\geq 0,
\end{equation}%
\begin{equation}
\text{DEC}\Leftrightarrow \rho -p=\beta +\alpha 2^{n+1}3^{n}(2n+1)\left(
H^{2}\right) ^{n+1}(nq+n+q-2)\text{ and }\rho =\frac{\beta }{2}-\frac{\alpha 
}{2}6^{n+1}(2n+1)\left( H^{2}\right) ^{n+1}\geq 0,
\end{equation}%
\begin{equation}
\text{SEC}\Leftrightarrow \rho +3p-6\overset{.}{f}_{Q}H+f=\alpha \left(
-6^{n+1}\right) (n+1)(q-1)\left( H^{2}\right) ^{n+1}\geq 0.
\end{equation}
\end{widetext}

\begin{figure}[tbp]
\includegraphics[scale=0.6]{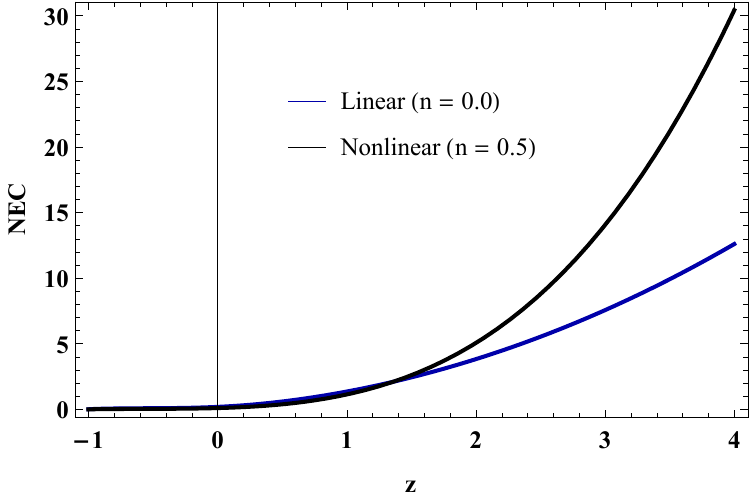}
\caption{Behavior of the NEC condition vs redshift.}
\label{FNEC}
\end{figure}

\begin{figure}[tbp]
\includegraphics[scale=0.6]{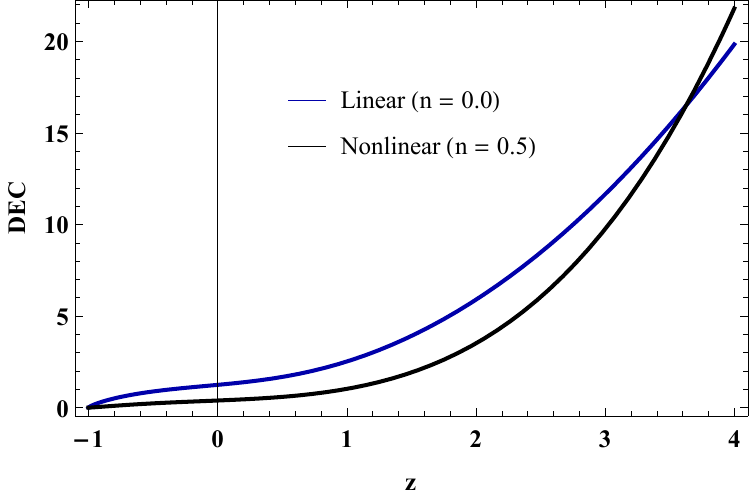}
\caption{Behavior of the DEC condition vs redshift.}
\label{FDEC}
\end{figure}

\begin{figure}[tbp]
\includegraphics[scale=0.6]{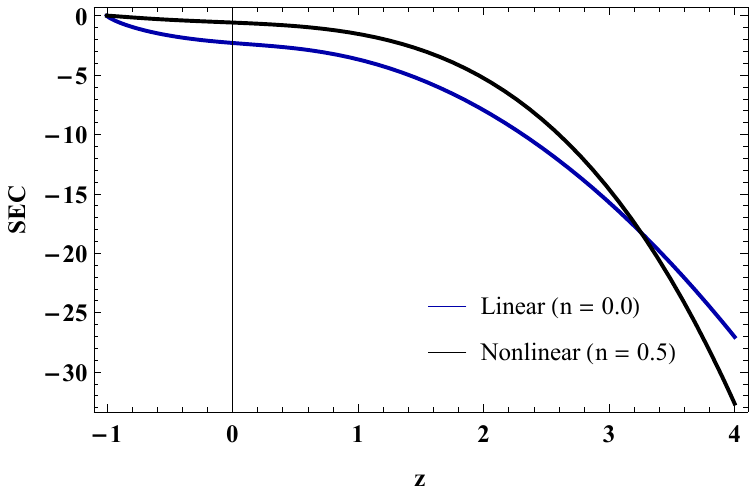}
\caption{Behavior of the SEC condition vs redshift.}
\label{FSEC}
\end{figure}

From Fig. \ref{Frho} it is clear that the WEC condition (energy density)
exhibits positive behavior. In addition, from Figs. \ref{FNEC}, \ref{FDEC}
and \ref{FSEC} we found that NEC and DEC conditions are satisfied, while the
SEC condition is violated. As we said, violation of SEC condition leads to
an acceleration phase of the Universe. The results obtained in our
cosmological model are consistent with many papers in the literature. 

\section{Conclusion}

\label{sec5}

The problem of the accelerating expansion of the Universe is one of the
greatest mysteries of modern cosmology present time. Despite the most
accepted $\Lambda $CDM model today can provide a significant match between
theoretical predictions and observations, it however, lacks a convincing
interpretation of the nature of dark energy (DE) described by the
cosmological constant. This mystery motivated us in this work to suggest a
simple cosmological model for dark energy within the framework of modified
gravity theories such as $f\left( Q\right) $\ symmetric teleparallel
gravity, which describes the gravitational effects with a geometrical
quantity called the non-metricity scalar $Q$. We considered two models to
discuss the results, the linear and nonlinear model by assuming the function 
$f\left( Q\right) $ as $f(Q)=\alpha Q^{n+1}+\beta $, where $\alpha $, $\beta 
$ and $n$ are free model parameters. Also, it is important to mention that $%
n=0$ corresponds to $f(Q)=\alpha Q+\beta $ (linear model), while $n\neq 0$
corresponds to a case of the nonlinear model. We discussed some cosmographic
parameters such as the jerk, snap and lerk parameters by assuming the
divergence free parametric form of the deceleration parameter which gave a
behavior consistent with the observations. Then, we obtained the expressions
for cosmic energy density, isotropic pressure and EoS parameter for our
cosmological model. We found that for both models specifically $n=0$ (linear
model) and $n=0.5$ (nonlinear model), the cosmic energy density is a
positive and increasing function in terms of the redshift, while the
isotropic pressure is negative in the present and the future. To find out
the dynamical nature of the model, we have discussed the $Om$ diagnostic and
find a negative slope of the latter indicating quintessence-like behavior. Further, by analyzing the behavior of the EoS parameter, we found that one
can get the quintessence model like behavior without introducing any DE
component or exotic fluid into the matter part. Thus, in our analysis cosmic
acceleration can only be explained by a geometrical generalization of
general relativity. Finally, we discussed all the energy conditions (WEC,
NEC, DEC and SEC) which were all fulfilled except for the SEC condition
which was violated and this is fine in the course of our analysis. This work
powerfully motivates us to look more at the new $f\left( Q\right) $\
symmetric teleparallel gravity.

\textbf{Data availability} There are no new data associated with this
article.

\textbf{Declaration of competing interest} The authors declare that they
have no known competing financial interests or personal relationships that
could have appeared to influence the work reported in this paper.\newline

%%%%%%%%%%%%%%%%%%%%%%%%%%%%%%%%%%%%%%%%%%%%%%%%%%%%%%%%%%%%%%%%%%%%%%%%%%%
%%%%%%%%%%%%%%%%%%%%%%%%%%%%%%%%%%%%%%%%%%%%%%%%%%%%%%%%%%%%%%%%%%%%%%%%%%%

\end{document}